\documentclass[a4paper]{jpconf}

\usepackage{graphicx}
\usepackage{amsmath,amsthm,latexsym,amssymb,amsfonts,url}
\usepackage{bbm}
\usepackage[usenames,dvipsnames]{color}
\usepackage{hyperref}

\hypersetup{
pdfkeywords={spinfoam, spin foam, loop quantum gravity, semiclassical limit, continuum limit, Regge calculus, Regge gravity, general relativity}, 
colorlinks=true,         
linkcolor=red,          
citecolor=Violet,        
urlcolor=Violet,      
}

\begin{document}

\newcommand{\mc}[1]{\mathcal{#1}}
\newcommand{\mbb}[1]{\mathbbm{#1}}
\newcommand{\ket}[1]{|#1\rangle}
\newcommand{\varket}[1]{[#1\rangle}
\newcommand{\bra}[1]{\langle#1|}
\newcommand{\comment}[1]{}

\newcommand{\im}[1]{\text{Im}\,#1}
\newcommand{\re}[1]{\text{Re}\,#1}

\title{Einstein-Regge equations in spinfoams\footnote{Based on a talk by the author at the conference {\it Loops '11} in Madrid on May 24th 2011 \cite{talkLoops11}.}}

\author{Claudio Perini}

\address{Institute for Gravitation and the Cosmos, Physics Department, Penn State, University Park, PA 16802-6300, USA}

\ead{claude.perin@gmail.com}

\begin{abstract}
We consider spinfoam quantum gravity on a spacetime decomposition with many 4-simplices, in the double scaling limit in which the Immirzi parameter $\gamma$ is sent to zero (flipped limit) and the physical area in Planck units ($\gamma$ times the spin quantum number $j$) is kept constant. We show that the quantum amplitude takes the form of a Regge-like path integral and enforces Einstein equations in the semiclassical regime. In addition to quantum corrections which vanish when the Planck constant goes to zero, we find new corrections due to the discreteness of geometric spectra which is controlled by the Immirzi parameter.
\end{abstract}

\section{Introduction}
Spinfoams \cite{Baez:1997zt,Reisenberger:2000fy} are a tentative covariant quantization of general relativity. They provide transition amplitudes between quantum states of 3-geometry, in the form of a Misner-Hawking sum over virtual geometries. In the so-called `new models' \cite{Engle:2007wy,Freidel:2007py} considered here, these virtual states are the ones of canonical loop quantum gravity, the $SU(2)$ spin-network states, a remarkable feature that promotes the spinfoam framework to a tentative path integral representation of Hamiltonian loop quantum gravity.

A major open problem is to understand how the semiclassical expansion of the spinfoam Feynman path integral reproduces classical Einstein equations. We expect something similar to the formal semiclassical approximation of the gravitational path integral
\begin{align}\label{whatweexpect}
\int Dg_{\mu\nu}e^{\frac{i}{\hbar}S_{EH}(g_{\mu\nu})}\sim e^{\frac{i}{\hbar}S_{EH}(g^0_{\mu\nu})}
\end{align}
in the classical limit $\hbar\rightarrow 0$, where $S_{EH}$ is the Einstein-Hilbert action for general relativity, evaluated on the r.h.s. on the classical solution $g^0_{\mu\nu}$ determined by the boundary conditions on the metric field $g_{\mu\nu}$. In \cite{Magliaro:2011dz,Magliaro:2011zz} the author (in collaboration with E. Magliaro) have proposed a tentative solution to this problem that has a compelling interpretation in terms of the interplay between \emph{two} physical scales in loop quantum gravity: the Planck length and the minimal physical length of loop gravity. To simplify the presentation, we work here in Euclidean signature.
\section{Spinfoam amplitudes: statement of the semiclassical regime}
A dual simplicial 2-complex $\sigma$ is a collection of vertices, faces, and edges which are dual to 4-simplices, triangles and tetrahedra respectively. The 2-complex is decorated in the following way. For each face $f$ there is an associated discrete area $a_f=\gamma\hbar\,j_f(j_f+1)$, with $j_f$ an integer spin. Faces are oriented and bounded by a cycle of edges $e$. The positive numer $\gamma$ is the Barbero-Immirzi parameter. Here we consider $0<\gamma< 1$. Each edge bounding a face has a source vertex $s(e)$ and a target vertex $t(e)$, where source/target is relative to the orientation of the face. To each edge let us associate $SU(2)$ elements $n_{ef}$ ($f$ runs over the faces meeting at the edge $e$) and two source/target $Spin(4)\simeq SU(2)\times SU(2)\sim SO(4)$ gauge group variables $g_{e,s(e)}$, $g_{e,t(e)}$. The variables $n_{ef}$ can also be interpreted as unit vectors $\vec n_{ef}$ in $\mathbbm R^3$, up to a phase ambiguity, by saying that $n$ is a rotation that brings a reference direction to the direction of $\vec n$.

The spinfoam amplitude \cite{Engle:2007wy,Freidel:2007py} for a 2-complex $\sigma$ without matter is given by
\begin{align}
Z(\sigma)=\sum_{\{a_f\}}\int dg_{ve}\int d\vec n_{ef}\, \exp{\Big(\frac{S}{\hbar}+\frac{S'}{\gamma\hbar}\Big)},
\label{amplitudejn}
\end{align}
where $S$ and $S'$ are gauge-invariant actions. We restrict our analysis to the nondegenerate part of the integral (the 4-volume of any 4-simplex cannot be arbitrarily small, that is smaller than a fixed $\delta>0$) as done in \cite{Conrady:2008mk}.

Now we state the semiclassical regime we are interested in as follows. We introduce two physical scales: the first one is the length scale $\lambda_{QG}$ of \emph{quantum gravity}, that can be identified with the Planck length, the second one is the scale $\lambda_{LQG}$ of \emph{loop quantum geometry}, that is the scale where we can `see' the discreteness of spacetime, according to the predictions of loop quantum gravity. We have
\begin{align}
l_{QG}\equiv l_P,\quad 
l_{LQG}\equiv\sqrt{\gamma}\,l_P.
\end{align}
The semiclassical regime of the spinfoam path-integral we are looking for is a two-scale problem expressed by the following relation:
\begin{align}
l\gg l_{QG}\gg l_{LQG},
\end{align}
where $l$ is the typical linear scale of each 4-simplex in the spacetime triangulation. Formally, we study this regime by letting $\hbar$ go to zero and at the same time letting $\gamma$ be small. This is done in a two-step stationary phase approximation as explained in the following.

\section{The flipped $\gamma\rightarrow 0$ limit}
The first step is the flipped limit expansion $\gamma\rightarrow0$, small Immirzi parameter, with physical area $a$ kept constant (`flipped' is named after the `flipped' spinfoam model \cite{Engle:2007qf} following a suggestion by C. Rovelli). In other words, restoring the spin quantum numbers, we consider large spins in such a way that
\begin{align}\label{limitconsidered}
j\rightarrow\infty,\;\gamma\rightarrow 0,\;j\gamma=const.
\end{align}
and $\gamma j$ is the macroscopic physical area in Planck area units. Notice that the Immirzi parameter controls the spacetime discreteness. In particular it controls the area gap and the spacing between area eigenvalues, thus the limit \eqref{limitconsidered} is a continuum limit for the area operator. 

Because the coefficient of $S'$ in \eqref{amplitudejn} is large, we use the extended stationary phase method to study the `flipped' regime and compute the main contribution to the path integral. The action $S'$ is complex with nonpositive real part $\text{Re}\,S'\leq 0$, so we expect that the main contribution can come only from the configurations such that $\text{Re}\,S'=0$. Splitting the group $SO(4)$ in selfdual and antiselfdual parts ($+$ and $-$), this condition is equivalent to the glueing equations
\begin{align}
g^+_{ev}\vec n_{ef}=-g^+_{e'v}\vec n_{e'f}\label{critical+}\\
g^-_{ev}\vec n_{ef}=-g^-_{e'v}\vec n_{e'f}
\label{critical-}
\end{align}
where $e,e'$ are adjacent edges in the face $f$, sharing the vertex $v$. Furthermore, we require that the critical points are stationary points, because otherwise the rapidly oscillating phase would suppress the amplitude. Varying $S'$ with respect to $SO(4)$ group variables, and evaluating at a critical point we get the closure condition
\begin{align}
\left.\delta_{g_{ev}} S'\right|_{crit.}=0\;\longrightarrow\;\sum_{f\in e}a_{f}\vec n_{ef}=0,
\end{align}
namely the closure relation for the tetrahedron dual to the edge $e$. The variations with respect to the unit vectors $\vec n_{ef}$ and with respect to the areas $a_f$ do not impose further restrictions, once the previous conditions are imposed.

It can be shown (see \cite{Magliaro:2011zz} and also the more recent \cite{Han:2011rf}) that the glueing and closure equations imply that a nondegenerate critical configuration $(j_f,\vec n_{ef},g_{ev})$ describe a 4-dimensional spacetime Regge triangulation. More precisely, the equivalence classes of parity-related nondegenerate critical points are in one-to-one correspondence with Regge triangulations. The areas of the triangulation are given by $a_f$, the 3-dimensional normals to the faces of the tetrahedra are specified by $\vec n_{ef}$, and the Levi-Civita connection is coded in the holonomies $g_{e,v}$. The simplicial metric tensor can be built out of these on-shell variables.

Curvature is distributional and concentrated on triangles. Moreover, the evaluation of the full action at a nondegenerate critical point with globally consistent parity gives the Regge action
\begin{align}
 S+\frac{1}{\gamma}S'|_{crit.}=i S_{\rm Regge}=i \sum_f a_f \Theta_f
\label{Sfonshell}
\end{align}
of general relativity, where the deficit angle $\Theta_f$ codes the curvature concentrated on the triangle dual to the face $f$. The other parity-related, possibly non globally consistent critical points yield a generalization of the Regge action (see also \cite{Barrett:1993db}). Each of them contibutes to the small $\gamma$ approximation of the path integral. In general this is very difficult to compute explicitly and depends on the simplicial 2-complex chosen, but we expect it to take the form of a sum over the set $\mathcal C$ of critical points, with some induced measure $\mu$. The result is
\begin{align}\label{previousintegral}
Z(\sigma)= \int_{\mathcal C} d\mu(a_f,n_{ef},g_{ev}) e^{i S_{\rm Regge}}+\rm{p.r.}+\mathcal O(\gamma)
\end{align}
where p.r. are the parity-related terms, and $\mathcal O(\gamma)$ denotes the $\gamma$-corrections to the partition function that come from the next orders of the `flipped' asymptotic approximation. 

Thus the spinfoam amplitude reduces to a Regge-like quantum gravity path integral \cite{Rocek:1982fr} given by formula \eqref{previousintegral}, up to $\gamma$-corrections. The latter take into account the effect of quantum discreteness of volumes and areas of loop quantum gravity. This is a remarkable feature of the `flipped' limit $\gamma\rightarrow 0$. The result resonates with the recent computations of the spinfoam graviton propagator \cite{Bianchi:2009ri}. In particular, the leading order graviton propagator $G(x,y)$ in the $\hbar$ expansion shows the same kind of $\gamma$-corrections,
\begin{align}\label{propagator}
G(x,y)= \frac{R+\gamma X+\gamma^2 Y}{|x-y|^2}+\hbar\text{-corr.}
\end{align}
and only in the limit $\gamma\rightarrow 0$ with physical area kept constant the 2-point function matches with the one computed in Regge gravity (the term proportional to the matrix $R$ in \eqref{propagator}). It also matches with a suitable projection of the standard propagator of perturbative gravity on flat background. The $\gamma$-corrections in \eqref{propagator} were also shown to break the parity symmetry of the propagator. Interestingly, similar corrections were found by Bojowald in the cosmological context \cite{Bojowald:2001ep}. 
\section{The semiclassical $\hbar\rightarrow 0$ limit}
Starting from \eqref{previousintegral}, we can parametrize the integral using length variables. Indeed given that the nondegenerate critical points correspond to Regge triangulations, there must exist an assignment of lengths $l_s$ to the sides $s$ of the triangles such that the areas $a_f$ coincide with the areas computed out of the lengths: $a_f=a_f(l_s)$. Let us parametrize the set of critical points $\mathcal{C}$ with the side lengths $l_s$ of the triangulation, and rewrite \eqref{previousintegral} as
\begin{align}\label{intlengths}
Z(\sigma)= \int d\tilde\mu(l_s) e^{\frac{i}{\hbar} S_R(l_s)}+{\rm p.r.}+\mathcal O(\gamma),
\end{align}
where the Regge action is now viewed as a function of the lengths.

Now we perform the second step of our semiclassical expansion which is the usual classical limit obtained formally in the asymptotic regime $\hbar\rightarrow 0$, as in a WKB expansion. Let us concentrate on the first term in \eqref{intlengths}. In the classical limit the stationary lengths of the Regge action dominate the integral. Using also the Schlafli identity, which tells us that the variation of the deficit angles do not contribute to the total variation of the action, these are the solutions of the Regge equations \cite{Regge:1961px}
\begin{align}\label{Ricciflat}
\sum_{f}\frac{\partial a_f}{\partial l_s}\Theta_f=0,
\end{align}
a simplicial version of the Einstein equations in vacuum, $R_{\mu\nu}=0$, namely of the vanishing of the Ricci tensor. The simplicial Einstein equations impose a relation between the deficit angles of different faces. Notice that there may be semiclassical contributions from the parity-related terms, for which a geometric interpretation is less transparent to us. Suppose the physical problem we are studying is such that there is only one solution of the Regge equations compatible with the boundary data; then by the stationary phase method the path integral \eqref{intlengths} goes like
\begin{align}
Z(\sigma)\sim e^{i S_{\rm Regge}}+{\rm p.r.}+\mathcal O(\gamma),\quad\quad\hbar\rightarrow 0
\end{align}
where $S_{\rm Regge}$ is now evaluated on the solution of \eqref{Ricciflat}, and thus can be interpreted as the Hamilton-Jacobi function. The last equation is a realization of the semiclassical approximation $\eqref{whatweexpect}$ in spinfoams, and shows that in the two-scale limit considered here the Einstein-Regge equations correctly emerge. 

We leave the study of the degenerate configurations as an open problem.
\newline

\bibliography{biblioreggefromsf}
\bibliographystyle{iopart-num}

\end{document}